\documentclass{iopart}

\usepackage{epsfig}
\usepackage{natbib}
\usepackage{graphicx}
\usepackage{bbm}

\def\newblock{\hskip .11em plus .33em minus .07em}

\usepackage{epstopdf}
\newcommand{\beq}{\begin{equation}}
\newcommand{\eeq}{\end{equation}}
\newcommand{\lsi}{\,\raisebox{-0.13cm}{$\stackrel{\textstyle<}
{\textstyle\sim}$}\,}
\newcommand{\gsi}{\,\raisebox{-0.13cm}{$\stackrel{\textstyle>}
{\textstyle\sim}$}\,}
\newcommand{\be}{\begin{equation}}
\newcommand{\ee}{\end{equation}}

\begin{document}

\title{Generalized Maximum Likelihood Method for Ultrahigh Energy Cosmic Rays}

\author{Glennys R. Farrar}

\address{Center for Cosmology and Particle Physics,
Department of Physics\\ New York University, NY, NY 10003, USA\\
}


\begin{abstract}
The Maximum Likelihood Method is generalized to include effects important for UHECR applications.  The new approach can incorporate source distance constraints implied by the observed CR energy and can allow for energy uncertainties, possible deflection in magnetic fields, multiple source types, and a spectrum of CR composition.  It can be efficiently implemented and does not require the unphysical ``isotropic" assumption for unidentified sources.  The approach optimizes the utility of UHECR data to discriminate between source classes and can help constrain galactic and extragalactic magnetic fields.  Aspects of the method are directly applicable in other contexts, such as TeV gamma ray astrophysics.

\end{abstract}


\section{Introduction}

The Maximum Likelihood (ML) method allows for measured or modeled uncertainties in a dataset to be incorporated in an analysis.  It allows data with different intrinsic resolution to be combined, and makes most efficient use of all available information.  Most importantly, it avoids the disagreeable feature of binned analyses on small datasets: fluctuations in the conclusion depending on exactly where cuts are made and thus which events are included.  The basic idea is to associate with each event in the dataset, a probability density for an event to be found with the observed characteristics (e.g., energy, arrival direction and possibly other properties such as composition indicators), for a given hypothesis as to how the observed UHECRs were produced and propagated to the detector: the type of sources, deflection properties of intervening magnetic fields, etc.  The production hypothesis is characterized by some assumptions (e.g., that all UHECRs are protons -- thus specifying the energy loss process -- and that UHECR sources reside in galaxies found in certain catalogs) and a small set of parameters (e.g., the number of UHECRs which originate in each source catalog, the mean-square magnetic deflection per unit propagation distance, and perhaps a luminosity threshold for the source).  The likelihood of finding the observed data-set is proportional to the product of the probability densities of the events making it up, and the most probable value of the parameters -- under the given hypothesis -- are those that maximize the likelihood.

Up to now, applications of the maximum likelihood method to UHECR data\cite{HR:BLLacs, jfML07} have considered only arrival direction correlations and angular resolution.  Energy resolution was not included and the constraint on probable source distance from energy losses during propagation (referred to collectively below as the GZK constraint, for short) was implemented only crudely, by imposing a hard cut on source distance.  However these are crucial features of the data and the ML method needs to be generalized to incorporate them.  The framework to do so, and to allow for multiple source types, possible deflection in magnetic fields, and CR composition as well, is presented below.  A computational strategy for efficient implementation is presented.  

The approach optimizes the utility of a given UHECR dataset (which can come from different experiments) to discriminate between source classes and can help constrain galactic and extragalactic magnetic fields.  The method is presented in terms of UHECRs but it can also be used to good effect in simpler contexts, such as TeV gamma ray astrophysics.  

\section{Maximum Likelihood Method} 
\label{sec:ML}

Since UHECRs may be produced by several different mechanisms (e.g., the jets of powerful radio galaxies\cite{biermannStrittmatter87}, Gamma Ray Bursts\cite{waxmanUHECR}, young magnetars\cite{aronsUHECR}, and Giant AGN Flares\cite{fg08}), the analysis must allow for multiple source catalogs labeled $k$.  The total number of catalogs being tested is $N_{\rm cat}$ and there are $N_k$ galaxies in the $k$th catalog.  Let $\nu_k$ be the (a priori, unknown) number of CRs coming from objects in the $k$th catalog and $\nu_{\rm tot}$ the total number of UHECRs in the dataset, $\Sigma_k \nu_k = \nu_{\rm tot}$.   

Given the angular resolution for the $i$th event -- deferring discussion of magnetic deflection until section \ref{sec:magdef} to simplify this introduction -- one can write the probability density to find that event at any position in the sky if it were produced by some source candidate (labeled $j$).  Call that probability density function $Q_i(x_i -s_j,\sigma_i,\, ...)$, where $\sigma_i$ characterizes the angular resolution of the event, $x_i$ is its arrival direction, and $s_j$ the direction of the source.  Unless deflection is large, flat geometry is adequate and $x, \,s_j,\, \sigma$ are 2D vectors in the plane of the sky;  for brevity, flat geometry is adopted and 2D vector notation is suppressed except where essential.   The ellipses in $Q(x-s_j,\sigma,...)$ denote information that is relevant when magnetic deflection is taken into account, deferred to section \ref{sec:magdef} below; $Q$ has unit normalization 
\be
\int d^2 x ~ Q(x,\sigma,\, ...) = 1.
\ee
For the case of no magnetic deflection and isotropic Gaussian resolution,
\be
\label{Q}
Q(x_i -s_j,\sigma) = \frac{1}{2 \pi \sigma^2} {\rm exp} \left( - \frac{(x_i -s_j)^2}{2 \sigma^2} \right);
\ee
for this distribution, a cone of radius $1.5 \sigma$ contains 68\% of the events.  

For each event, one needs the relative probability for it to come from a source at different redshifts.   Due to GZK energy losses, this depends strongly on the observed energy $E$ above about 50 EeV.  It also depends on the assumed CR production spectrum at the source, since a harder spectrum and higher maximum energy mitigate the energy attenuation to some extent.  A natural assumption, unless one has a large enough data-set to potentially have multiple events from individual sources or there is external information, is that every candidate source in a given catalog produces CRs with the same spectrum, apart from overall normalization.  It is more convenient to work with $\epsilon \equiv ln\, E$, and we denote the probability density to observe a CR in the range $\epsilon, \, \epsilon + d \epsilon$ from a source at distance $z, \, z + dz$ by $F(\epsilon, z, \lambda)$
;
 $\lambda$ characterizes the energy resolution.  For a fixed energy of the observed cosmic ray, $F(\epsilon, z_j, \lambda)/F(\epsilon, z_{j'}, \lambda)$ is the relative likelihood that the CR comes from a source at $z_j$ compared to one at $z_{j'}$. 

For example, we could take the source spectrum to be a power-law, $E^{-p}$, with $\epsilon_{\rm min} \leq \epsilon \leq \epsilon_{\rm max}$ and $p = 2$ or 2.5.  The results of the analysis will depend on the choice of $p$ and $\epsilon_{\rm max}$.  The analysis should be done for several model spectra (here, values of $p$ and $\epsilon_{\rm max}$) to estimate the systematic uncertainty in the ML method due to the spectrum.  It may also be possible to treat $p$ and $\epsilon_{\rm max}$ as parameters to be determined in the analysis.  As long as $\epsilon_{\rm min}$ is low enough that it includes the minimum true arrival energy given the energy resolution and $E_{\rm thresh}$, and low enough that energy losses in propagation do not move CRs from the most distant source out of the range $\epsilon^\prime \geq \epsilon_{\rm min} $, the value of $\epsilon_{\rm min} $ does not affect the result.  The absolute normalization of $F$ does not matter, but must be the same for each source.  

We will denote the true energy at Earth by $E^\prime$ and $\epsilon^\prime \equiv ln\, E^\prime$ and define $G(\epsilon, z)$ to be the probability density for a CR reaching Earth from a source at distance $z, \, z + dz$ to have a true $ln$ energy in the range $\epsilon^\prime, \, \epsilon^\prime + d \epsilon^\prime$, given the assumed initial spectrum.  
If the energy uncertainty is just a log-normal energy resolution of width $\lambda$,  then
\be
\label{F}
F(\epsilon, z,  \lambda) = \int d \epsilon^\prime \, G(\epsilon^\prime, z)  \frac{1}{\sqrt{2 \pi}\,  \lambda}\, {\rm exp} \left( - \frac{(\epsilon^\prime -\epsilon)^2}{2 \lambda^2} \right).
\ee
In order to appropriately call them probability densities, one can choose
\be
\int_{\epsilon_{\rm min}}^{\epsilon_{\rm max}} d \epsilon^\prime \, G(\epsilon^\prime ,z) = 1; ~~~~~  \int_{\epsilon_{\rm min}}^{\epsilon_{\rm max}} d \epsilon \, F(\epsilon,z, \lambda) = 1.  
\ee

Note that one can prepare a lookup table for $F(\epsilon_i,  z_j, \lambda_i)$ where $\epsilon_i,\, \lambda_i$ are the $ln$ energy and its resolution, of the $i$th UHECR, and the $z_j$'s are the source redshifts.  Then, the generalized ML (GML) analysis runs just as fast as it would ignoring energy resolution.  If there is a known systematic overall energy shift for some set of events, those event energies would simply be shifted before doing the analysis.  If it is desired to infer the most likely value of the energy shift of a particular experiment, then the energy of each event from that experiment would be multiplied by a factor, call it $\kappa_{\rm exp}$, and $\kappa_{\rm exp}$ would be an additional parameter to be determined by maximizing the likelihood as discussed below; events from different experiments can be assigned different values of $\kappa_{\rm exp}$ and these values determined independently if the data is adequate.

With the above functions defined, the probability density for a given UHECR event to be found having the observed parameters is just the sum of the probability density that it comes from each of the individual sources in the catalogs, which can be written as
\be
\label{P}
\mathcal{P}_i(x_i, \epsilon) = \Sigma_{k = 1}^{N_{\rm cat}} \, \left ( \frac{\nu_k}{\nu_{\rm tot}} \right ) ~ \Sigma_{j = 1}^{N_k} ~ w^{(k)}_j \, Q(x_i-s_j,\sigma_i,...)\, F(\epsilon_i,z_j, \lambda_i),
\ee
where $w^{(k)}_j$ is the weight attributed to the $j$th source in the $k$th catalog.  Each catalog must have the same total weight, so that the relative weight of the catalogs is just $\nu_k/\nu_{\rm tot}$. This we enforce by the normalization condition
\be
\label{wtnorm}
 \Sigma_{j = 1}^{N_k} \, w^{(k)}_j = 1.
 \ee
In the absence of any specific knowledge to the contrary, each source has equal a priori probability and thus equal weight.  However it is clear that the probability of a particular source contributing to the dataset is proportional to the relative exposure, $\eta_j$, of the UHECR dataset at position $s_j$.  Furthermore, the flux from a source falls as $z^{-2}$, so in the absence of more information about individual sources, $w^{(k)}_j$ would be taken proportional to $\eta_j \, z_j^{-2}$.  Additional information about the sources can be included in $w^{(k)}_j$ if available. For instance, sources in the IRAS catalog might be weighted with their infrared luminosity.  Thus a typical choice for weights might be
\be
\label{typwt}
w^{(k)}_j  =  \eta_j \, f_j^{(k)} \left(  \Sigma_{j = 1}^{N_k}\, \eta_j \, f_j^{(k)} \right)^{-1},
\ee
where $f_j^{(k)}$ is the measured flux of the $j$th element of the $k$th catalog.  Note that if the number of objects in a catalog is large, the normalization condition (\ref{wtnorm}) implies that the weights associated with individual objects in that catalog are correspondingly reduced, as they should be.  Multiple source candidates may fall within the angular domain of a given UHECR, in which case all of them would contribute according to their weights.  

Since the candidate source catalogs cannot in general be assumed to actually contain all the sources of the events, a ``background" option must be included in the $k$ catalogs in eqn (\ref{P}).  Traditionally this has been taken to be an isotropic distribution, but on astrophysical grounds one expects the sources of UHECRs to be concentrated in regions of higher matter density.  Thus a better treatment would use the 3D density distribution derived from a complete, unbiased catalog of galaxies.  In practice obtaining such a density distribution is not trivial, due to incompleteness in the Galactic plane for optical surveys and the problem of too-low sampling for spatially more complete catalogs like HIPASS as used in ref. \cite{Ghisellini08}.  

The traditional method of allowing for ``background" adds an isotropic term to eqn (\ref{P})
\be
 ... \, + \frac{(\nu_{\rm tot} - \Sigma_{k = 1}^{N_{cat}}\, \nu_k)}{\nu_{\rm tot}} \, R(x),
\ee
where $R(x)$ is the relative acceptance in the direction $x$ (normalized so that 
$ \int d^2 x \, R(x) = 1$).  This approach distorts the results of the analysis and should only be used for qualitative or preliminary studies.  For instance, in a study for which UHECRs are compared to known BL Lacs versus an isotropic source distribution, the analysis assuming isotropic background will return a larger number of BL Lac as sources than if the background were taken to be the galaxy distribution, even if the true UHECR sources are some other type of galaxy and not BL Lacs, simply because BL Lacs are clustered with galaxies. 

A better way to allow for the possibility that the chosen source catalog(s) do not contain all the sources of the UHECRs in the dataset, is to include a ``background catalog" in the list of $k$ possible source catalogs.  The background catalog should be volume-limited and unbiased.  The criterion for adequate sampling density for a background catalog is that some source in the catalog should be within an angular separation $\leq \sigma$ of each UHECR, and within the redshift separation over which $F(\epsilon,  \lambda, z)$ is slowly varying out to the largest contributing redshift, given the energy threshold of the UHECR dataset.  If that condition cannot be satisfied with a given all-sky catalog, then the choices are to use a catalog such as 2MASS Redshift Survey (2MRS) and use its mask to apply a corresponding hard cut to the domain from which the CR dataset is taken, increase the energy threshold of the UHECR dataset to reduce the relevant range of $z$ so it is low enough that the condition can be met, or a combination of the above.

Note that it is in general incorrect to use an isotropic distribution to ``complete" a catalog in either redshift or arrival direction.  This applies to either the background catalog or to a source-candidate catalog.  As shown in \cite{jfML07}, the ML method automatically imposes the condition that the global distribution of matches to each catalog should represent the (weighted) global distribution of elements of each catalog.  Thus a non-realistic global distribution for a catalog also distorts the $\nu_k$'s obtained.  This is why it is for instance not correct in principle to fill in a ``background" catalog like 2MRS with an isotropic distribution in the region of the mask, or complete a source catalog like the Veron-Cetty Veron one, with an isotropic distribution for distance $\geq 100$ Mpc.   On the other hand, using an isotropic distribution as an approximate completion of a global under-sampling for the background catalog may be acceptable; such a procedure needs further study with simulations.

\section{Maximizing the Likelihood}

The next step is to introduce a likelihood measure for the entire dataset, and maximize with respect to the unknown parameters, to find the most likely values of these parameters. In the simplest case, these are simply the $k-1$ independent $\nu_k$ values. It is computationally desirable to work with the logarithm of the product of the $\mathcal{P}_i$'s for each of the individual events:
\be
\label{LM}
LM \equiv \Sigma_i^{\nu_{\rm tot}}\, ln \mathcal{P}_i.
\ee
Assuming Gaussian angular resolution, each $Q$ falls off very rapidly.  For instance restricting to source candidates within a radius $3 \sigma$ of a cosmic ray captures all but 1\% of the likely sources.  Furthermore, source candidates that are left out have $Q \lsi 0.01 \, Q_{\rm peak}$ compared to a perfectly aligned source with the given angular resolution.  With such a restriction on the angular region included for each cosmic ray, computation of the $LM$ of a given dataset for fixed values of the unknown parameters is straightforward, and not very time-consuming even if the background catalog is large.  Standard methods for multi-parameter maximization can be used to find the parameter values that maximize $LM$.  The test for whether a given catalog is relevant or not, is to redo the calculation successively eliminating the catalog with the smallest $\nu_k$.  If the $LM$ changes by more than one unit, that catalog is a significant contributor to the signal and should be kept.   The error ellipsoid in the multidimensional parameter space can be determined by finding the parameter values that give $LM$ values one, two, ... units lower than the maximum.   Note that the parameters $\nu_k$ are not in general integers -- they are the mean in the sense of a Poisson distribution of the number of events contributed by $k$th source class.

The analysis can be extended to larger number of unknown parameters by first determining the essential catalogs and values of key parameters.  Then, more parameters can be added, maximizing $LM$ with respect to all parameters but now keeping the domain of variation of the original parameters small.  

\section{Magnetic deflections}
\label{sec:magdef}

Charged UHECRs are deflected during their propagation between source and Earth, but extragalactic and Galactic magnetic fields (EGMF and GMF) are poorly constrained.  It may be possible to use UHECR correlations to constrain these fields and at the same time improve the  source analysis discussed above.  That is where the ellipses in $Q(x-s_j,\sigma,\, ...)$ come in.  When magnetic deflections are included, $Q$ depends not only on $x-s_j$ and $\sigma$, but also on $E$, the absolute arrival direction $x$, and additional parameters characterizing the magnetic deflections.  To what extent it will be possible to both identify the sources and constrain the magnetic deflections depends on how rare and distinctive UHECR sources are and whether there are multiple source types.  If sources are sparse and clearly identifiable, e.g., powerful AGNs, then multiple events from single sources will both strengthen the identification of their source and constrain the magnetic deflection and dispersion along that line-of-sight.  However if virtually every galaxy can produce UHECRs (e.g., at the birth of a magnetar) but such events are rare and the mean number of events from a single source is $\ll 1$, it will be difficult to use UHECRs to map the GMF.

In the limit of small deflections, the effect of a large scale magnetic field can be described in terms of a displacement function, $\vec{\beta}(x)$, giving the angular displacement due to the Lorentz force per unit CR rigidity (energy/$Z$) experienced as a CR traverses the field. 
For large scale fields and high enough CR energy, the trajectory in the Galaxy can be approximated as almost linear and $\vec{\beta}$ varies slowly with arrival direction.  Measuring UHECR energies in EeV, distances in kpc, angular separations in degrees, and magnetic fields in $\mu$G, the deflection of a CR traversing a distance $L$ through the GMF along the line of sight $\vec{l}$ is $\vec{\beta} \, Z \, E$ where
\be
\label{magdef}
\vec{\beta} = 630 \frac{1}{L} \int \hat{l} \times \vec{B}\, d\, l .
\ee
Under these assumptions, the coherent displacement of events from a single source can be described by a common $\vec{\beta}$.  The effect of magnetic deflection is incorporated in the GML analysis by changing $\vec{x_i}$ to $\vec{x_i} - \vec{\beta}(x_i) \, Z_i \, E_i$, so that $Q$ becomes a function of energy.  In practice, one would adopt some simple model of the GMF and maximize the $LM$ with respect to its parameters.  If the impact of energy resolution on arrival direction correlations is to be fully taken into account, the integral over $\epsilon$ must include $Q$ as well as $G$. 

In addition to coherent deflections by the GMF, propagation through turbulent random fields may be an important effect.  In the limit of sufficiently many small, randomly oriented deflections, the smearing of a cosmic ray from a pointlike continuous source due to random fields is described by a 2-d Gaussian probability distribution of width 
\begin{equation} 
\label{magsmear}
\sigma^2_B(E,D) \equiv \left( \frac{E^*}{E} \right)^2= \frac{ \langle Z^2 B^2 \lambda \rangle D}{9 E^2} ~~,
\end{equation}
where $D$ is the distance to the source and $\lambda$ is the characteristic length scale of the turbulent fields (expected to be $\sim$ Mpc for extragalactic but $\lsi 100$ pc for Galactic random fields).  The expression (\ref{magsmear}) is easily understood in the case of randomly oriented magnetic domains of size $\lambda$ and strength $B$, where the magnitude of the deflection with respect to the direction of motion in each domain is $ \delta \theta = \lambda/R_L$ and $R_L = 1.08 \, {\rm kpc} \frac{E_{\rm EeV}}{Z B_{\perp, \rm nG}}$.  The total number of deflections is $N = D/\lambda$ and the mean total deflection obeys $\Delta \theta^2 =N \delta \theta^2/3 $.  The factor 1/3 compared to the familiar random walk formula was derived in \cite{alcockH} for the case of x-ray scattering off dust, and arises because $\Delta \theta$ is the angle between the UHECR velocity vector and a vector pointing to the source, both of which change as the UHECR propagates.  Isotropy implies $<B_{\perp}^2>\, = \frac{2}{3} <B^2>$.  The general expression for $\langle B^2 \lambda \rangle$  as a weighted mean of $B^2 \lambda$ over the trajectory is given in \cite{waxmanME}.   Equation (\ref{magsmear}) is applicable if $\sqrt{N} \sim \sqrt{D/<\lambda>} \gg 1$.  If multiple events come from the same source, their trajectories must sample different magnetic domains for them to have independent deflections, requiring that the {\it difference} in their displacements must be larger than a characteristic domain size, i.e., 
\beq \label{cond}
\left | \frac{E_*}{E_i} - \frac{E_*}{E_{i'}} \right | D \gsi \langle \lambda \rangle ,
\eeq
for all events $i,~i'$ in the cluster under consideration.  

Since the experimental angular resolution of the $i$th event is approximately a symmetric 2-d Gaussian (of width $\sigma_{i}$),  the net result of Gaussian magnetic smearing is to increase the effective resolution of that event in computing $Q$ for the $j$th source, to $\sigma_{i, \rm eff} = \sqrt{\sigma_{i}^2 + \sigma_B(E_i,Z_i,D_j)^2}$.  For the simplest treatment in a correlation analysis between UHECRs and catalogs as discussed above, one could take 
\be
\label{beta}
\sigma^2_B(E_i,Z_i,D_j) \rightarrow \frac{ \beta_0 D_j}{ (E_i/Z_i)^2} ~~
\ee
with $\beta_0$ a single universal parameter to be determined by maximizing the $LM$.  Note that $\langle B^2 \lambda \rangle$ need not be constant for the description (\ref{beta}) to be adequate, as long as the distribution of random magnetic deflections is Gaussian; the inferred value of $\beta_0$ would be the mean of the distribution.  A more sophisticated treatment could be to assume the value of $\langle B^2 \lambda D \rangle$ for the $j$th source candidate is proportional to some measure of the energy density of intervening IGM, e.g., the integrated luminosity of background galaxies along the line-of-sight to $j$ -- this would correspond to the reasonable assumption that the energy density in random magnetic fields is proportional the the luminosity of the galaxy.  In this case, the proportionality constant could be determined by maximizing $LM$.  A virtue of the GML method is that as the data-set increases, more complex hypotheses for the EGMF can be used.

\section{Composition}
In order to carry out the magnetic deflection analysis one needs to know or make an assumption about the charges of the UHECRs.  As argued by \cite{allard&Composition08} it is likely that the highest energy cosmic rays have a pure or bimodal composition, due to the fragility of light and intermediate mass nuclei to photodisintegration during propagation.  Thus it is natural to assume that all events are protons or Fe.  However as UHECR reconstruction techniques improve, it will become possible to give individual UHECR events a probability distribution over primary particle type and hence charge, based on measured properties of the individual CR shower.  If this information is available, eqn (\ref{P}) is replaced by 
\be
\label{PZ}
\mathcal{P}_i(x_i, \epsilon) = \Sigma_{k = 1}^{N_{cat}} \, \frac{\nu_k}{\nu_{\rm tot}} ~ \Sigma_{j = 1}^{N_k} \, w^{(k)}_j \, p_{i,j}(x_i,\epsilon)
\ee
where now 
\be
p_{i,j}(x_i,\epsilon_i) =  \Sigma_Z ~\zeta^{(Z)}_i \, Q(x_i-s_j,\sigma_i,Z,...)\, F_Z(\epsilon_i,z_j, \lambda_i),
\ee
where $\zeta^{(Z)}_i$ is the probability for the $i$th event to have charge $Z$, normalized to
\be
 \Sigma_Z ~\zeta^{(Z)}_i = 1.
 \ee
The energy loss during propagation depends on the charge of the UHECR, so $F$ is generalized to $F_Z$.  Similarly, in the presence of magnetic deflection, $Q$ depends on $Z$ via eqns (\ref{magdef},\ref{magsmear}).  

Prior to the time that reconstruction methods have improved to the point that individual events can be assigned $\zeta^{(Z)}_i $ values reliably, one can use eqn (\ref{PZ}) to constrain the fraction of particles of charge $Z$ in the dataset.  For instance if the highest energy cosmic rays are assumed to be a mixture of protons and iron, one would take $\zeta^{(1)}_i = f_p$, $\zeta^{(26)}_i = 1-f_p$, all other $\zeta^{(Z)}_i =0$, and maximize the LM with respect to $f_p$.  Clearly, the success of such an effort depends on having a large high energy dataset and source catalogs that contain the sources of most of the events.

\section{Application to clusters of events}
The GML method developed here works whether or not there are multiple events from a single source.  However when a possible cluster is identified, as for instance the 4-5 events of the ``Ursa Major" cluster in the combined AGASA-HiRes dataset\cite{fBurstICRC07}, it is no longer necessary to correlate the members of the cluster to a candidate source, allowing for a self-contained analysis.  Rather than using eqn (\ref{P}) and maximizing the $LM$ with respect to the $\nu_k$'s for the different source catalogs, one introduces a source position $x_s$, redshift $z_s$, and number $\nu_s$ of events from that source.  Like the $\nu_k$'s in eqn (\ref{P}), $\nu_s$ is not in general an integer.  One maximizes the $LM$ to determine the most likely source position and redshift, and the most likely number of events coming from the source.  The alternative to coming from the source now being that they come from the background which as discussed above might be approximated by a catalog, $b$:  
\begin{eqnarray}
\mathcal{P}_i(x_i, \epsilon_i, x_s,z_s)  & = & \left ( \frac{\nu_s}{\nu_{\rm tot}} \right ) ~  \, Q(x_i-x_s,\sigma_i,...)\, F(\epsilon_i, z_s,\lambda_i) \nonumber \\
 && +   \left ( 1- \frac{\nu_s}{\nu_{\rm tot}} \right ) \Sigma_{j = 1}^{N_b} ~ w^{(b)}_j \, Q(x_i-s_j,\sigma_i,...)\, F(\epsilon_i,z_j, \lambda_i).
 \label{Pclus}
\end{eqnarray}

The $LM$ is formed in principle as in eqn (\ref{LM}), but the computational speed is greatly increased by separating the computation of the sum over UHECR events in eqn (\ref{LM}) into two parts, those which are near and those which are far from the cluster.  The $ ln \mathcal{P}_i$'s of UHECRs which are too far in angle to be in the cluster, have a negligible $Q$ value in the first term (proportional to $\nu_s$) in eqn (\ref{Pclus}).  Furthermore, the factor multiplying $( 1- \frac{\nu_s}{\nu_{\rm tot}} )$ in the second term of (\ref{Pclus}) (the sum on $j$) is independent of $x_s$ and $\nu_s$, the parameters to be varied.  Thus the contribution of distant UHECRs to $ \Sigma_i \, ln \mathcal{P}_i$ amounts to a piece that can be computed once and for all, plus $\nu_{\rm far}~ ln ( 1- \frac{\nu_s}{\nu_{\rm tot}} )$, where $\nu_{\rm far}$ is the number of distant events.  Thus the maximization of $LM$ over $x_s$ and $\nu_s$ is completely tractable computationally.  

The magnetic deflection and dispersion of the cluster events can be described by 3 parameters, $\vec{\beta}$ and $E^*$, for a total of 7 parameters including $x_s,\, z_s$ and $\nu_s$.  A simpler version of this analysis for the Ursa Major cluster, without the energy-distance constraint and using the traditional isotropic background, was reported in \cite{fBurstICRC07,gfclus}.  The significance of the cluster is similar for either 4 or 5 events: $\sim 0.2$\% probablity of being a chance fluctuation.  Even with only 4-5 candidate events in the cluster, the results for $\beta$ and $E^*$ are robust to variations in other parameters.  The coherent magnetic deflection is negligible and $\langle B^2 \lambda \rangle D  \approx 8 {\rm \, nG^2 \, Mpc^2}$ (assuming $Z=1$) \cite{fBurstICRC07}.  

The treatment of $F$ described in section \ref{sec:ML} above, using a power-law source distribution, applies only for a continuous source.  A point emphasized in \cite{fBurstICRC07} is that if a cluster of events comes from a bursting source, whose duration is short compared to the typical delay time from magnetic deflections, then the observed spectrum is peaked rather than power-law and this must be taken into consideration.  An analysis of the Ursa Major cluster in the spirit of the method outlined here, but applicable also for a bursting source, will be presented elsewhere.

\section{Discussion and Conclusions}
We have presented a general and practical way to maximize the utility of the ML approach for UHECR correlation studies.  Its greatest immediate value is in incorporating the source distance information following from the GZK energy loss phenomenon.  In first applications a simple composition could be assumed, but the method is general enough to allow for a distribution of possible charges of the UHECR and eventually a probability distribution for the charge of individual events when event-by-event composition indicators are improved.   The method can be applied directly to the simpler case of TeV gamma ray data, to include the effects of energy loss from $e^\pm$ production in interactions with diffuse background radiation.  A rescaling factor for the level of the background radiation in wavebands where it is poorly measured could be taken as a parameter to be determined by maximizing the LM.

At present, the published high quality data on UHECRs consists of 57 events from AGASA with energies above 40 EeV\cite{AGASAupdate}, 271 events from HiRes above 10 EeV (energy-ordering but not individual energies are available, G. Thompson private communication), and 27 events from Auger with energies above 56 EeV\cite{augerLongAGN}.  (The energy and angular resolution of earlier experiments is lower and more uncertain so they have only marginal utility in a ML analysis.)  With $\sim 400$ CRs whose typical resolution element is $\lsi 10$ sq-degrees and a total aperture containing $\sim 40$k sq-degrees, and future analyses most likely oriented to the higher energy range to get the benefit of the GZK horizon,  the dataset is small enough that the probability of two UHECRs from an isotropic dataset falling by chance within one characteristic angular resolution element is fairly small and the methods presented here should be quite powerful.  Simulations to test its effectiveness under different scenarios of source density and catalog completeness are needed.  
 
The Auger collaboration has reported a statistically signficant correlation between the highest energy UHECRs and Veron-Cetty Veron galaxies with $z < 0.018$ within $3.2^\circ$\cite{augerScience07,augerLongAGN}.    An initial ML study applying the generalized method proposed here to the 27 published events, might assume a pure proton composition and allow for an overall energy rescaling parameter and a single magnetic smearing parameter as in eqn (\ref{magsmear}).  This should have more discriminating power in a search for the sources of UHECRs, as a result of employing a more realistic background catalog and not requiring a hard cut in angular separation and $z_{\rm max}$ as in the original, binned Auger analysis.  Applied to the full Auger dataset including events with lower energy, one would expect that the number of events $\nu_k$ attributed to the $k$th candidate source catalog, and the significance of the correlation with it, should stay approximately constant as the energy threshold is lowered, with the number of events attributed to the ``background" catalog increasing as the dataset does. This is because the ``utility" of the added lower energy events to discriminate between source and background catalogs drops, as the number of candidate or background sources within a few-$\sigma$ cone about the UHECR direction becomes large with  increasing GZK horizon.  Thus the method should be insensitive to the energy threshold of the UHECR dataset, as long as the redshift-completeness of the source and background catalogs are comparable; otherwise, artifacts can be introduced.  Simulations to confirm this are underway.

The author acknowledges useful conversations with fellow members of the Pierre Auger Collaboration.  This research has been supported in part by NSF-PHY-0701451.

\bibliographystyle{unsrt}

\end{document}